\documentclass[%
 reprint,%
 amssymb, amsmath,%
 aip,cha%
]{revtex4-1}

\usepackage{bm}%

\usepackage{graphicx}
\usepackage{natbib}

\begin{document}

\title{Magnetization reversal, damping properties and magnetic anisotropy of \textit{L}$1_0$ – ordered FeNi thin films}


\author{V. Thiruvengadam}
\affiliation{Laboratory for Nanomagnetism and Magnetic Materials (LNMM), School of Physical Sciences, National Institute of Science Education and Research (NISER), HBNI, P.O.- Bhimpur Padanpur, Via –Jatni, 752050, India}
\author{B. B. Singh}
\affiliation{Laboratory for Nanomagnetism and Magnetic Materials (LNMM), School of Physical Sciences, National Institute of Science Education and Research (NISER), HBNI, P.O.- Bhimpur Padanpur, Via –Jatni, 752050, India}
\author{T. Kojima}
\affiliation{Institute for Materials Research, Tohoku University, 2-1-1 Katahira, Aoba-ku, Sendai 980-8577, Japan}
\author{K. Takanashi}
\affiliation{Institute for Materials Research, Tohoku University, 2-1-1 Katahira, Aoba-ku, Sendai 980-8577, Japan}
\author{M. Mizuguchi}
\email{mizuguchi@imr.tohoku.ac.jp}
\affiliation{Institute for Materials Research, Tohoku University, 2-1-1 Katahira, Aoba-ku, Sendai 980-8577, Japan}
\author{S. Bedanta}%
\email{sbedanta@niser.ac.in}
\affiliation{Laboratory for Nanomagnetism and Magnetic Materials (LNMM), School of Physical Sciences, National Institute of Science Education and Research (NISER), HBNI, P.O.- Bhimpur Padanpur, Via –Jatni, 752050, India}%

\date{August 2019}%


\begin{abstract}
$\textit{L}1_0$ ordered magnetic alloys such as FePt, FePd, CoPt and FeNi are well known for their large magnetocrystalline anisotropy. Among these, $\textit{L}1_0$-FeNi alloy is economically viable material for magnetic recording media because it does not contain rare earth and noble elements. In this work, $\textit{L}1_0$-FeNi films with three different strengths of anisotropy were fabricated by varying the deposition process in molecular beam epitaxy system. We have investigated the magnetization reversal along with domain imaging via magneto optic Kerr effect based microscope. It is found that in all three samples, the magnetization reversal is happening via domain wall motion. Further ferromagnetic resonance (FMR) spectroscopy was performed to evaluate the damping constant and magnetic anisotropy. It was observed that the FeNi sample with moderate strength of anisotropy exhibits low value of damping constant $\sim$ $4.9 \times 10^{-3}$. In addition to this, it was found that the films possess a mixture of cubic and uniaxial anisotropies. 
 
\end{abstract}

\maketitle

In order to increase the storage density in magnetic recording media it requires reduction in the bit size \cite{bandic2008advances}. On the other hand, in ferromagnetic materials, the superparamagnetic (SPM) limit is inevitable at a critical radius, below which the magnetic moment is thermally unstable and become incapable of storing the information \cite{bedanta2008supermagnetism}. Therefore, magnetic material with high magnetocrystalline anisotropy is essential to overcome the SPM limit \cite{weller1999thermal}. In this context, the $\textit{L}1_0$ ordered magnetic alloys such as FePt, FePd, CoPt and FeNi have potential for ultra-high density magnetic recording media because of their large uniaxial magnetocrystalline anisotropy energy density $\sim$ 10$^{7}$ erg cm$^{-3}$. $\textit{L}1_0$ ordered alloy is a binary alloy system with face centered tetragonal (FCT) crystal structure where each constituent atomic layers are alternatively laminated along the direction of crystallographic c-axis \cite{kotsugi2013origin,goto2017synthesis}. In these materials the large anisotropy energy is due to their tetragonal symmetry of $\textit{L}1_0$ crystal structure \cite{klemmer2003combined}.

$\textit{L}1_0$-FeNi possess high values of saturation magnetization (1270 emu cm$^{-3}$), coercivity (4000 Oe), and uniaxial anisotropy energy density (1.3$\times$10$^{7}$ erg cm$^{-3}$) \cite{kojima2014fe,takanashi2017fabrication,mizuguchi2010characterization}. In addition it’s Curie temperature is quite high $\sim$ 550$^{\circ}$C and it exhibits excellent corrosion resistance \cite{kojima2014fe}. All these above properties make $\textit{L}1_0$-FeNi a promising material for fabricating information storage media and permanent magnets. Further, $\textit{L}1_0$ ordered FeNi alloy is free of noble as well as rare-earth elements. Also the constituent elements (i.e. Fe and Ni) are relatively inexpensive. Therefore $\textit{L}1_0$ ordered FeNi alloy is economically viable for commercial applications\cite{kotsugi2011determination,mibu2015local,kotsugi2013origin}. It should be noted that although, $\textit{L}1_0$ - FeNi possess high uniaxial anisotropy, shape anisotropy becomes dominant in thin films \cite{mibu2015local}. Previously the study of the order parameter (S) and Fe-Ni composition dependence on $\textit{K$_{u}$}$  has been reported \cite{kojima2014fe}. Previously, magnetic damping constants for $\textit{L}1_0$-FeNi and disordered FeNi have been studied employing three kinds of measurement methods\cite{ogiwara2013magnetization}. However, the effect of different anisotropy values of $\textit{L}1_0$ - FeNi thin films on evolution of their magnetic domain structures and damping has not been studied so far. Therefore, focus of this paper is to study the domain structures during magnetization reversal, anisotropy strength, damping properties of $\textit{L}1_0$ - FeNi thin films with different anisotropy values. 

\begin{figure*}[htb]
   \begin{center}
   \includegraphics[scale=0.6]{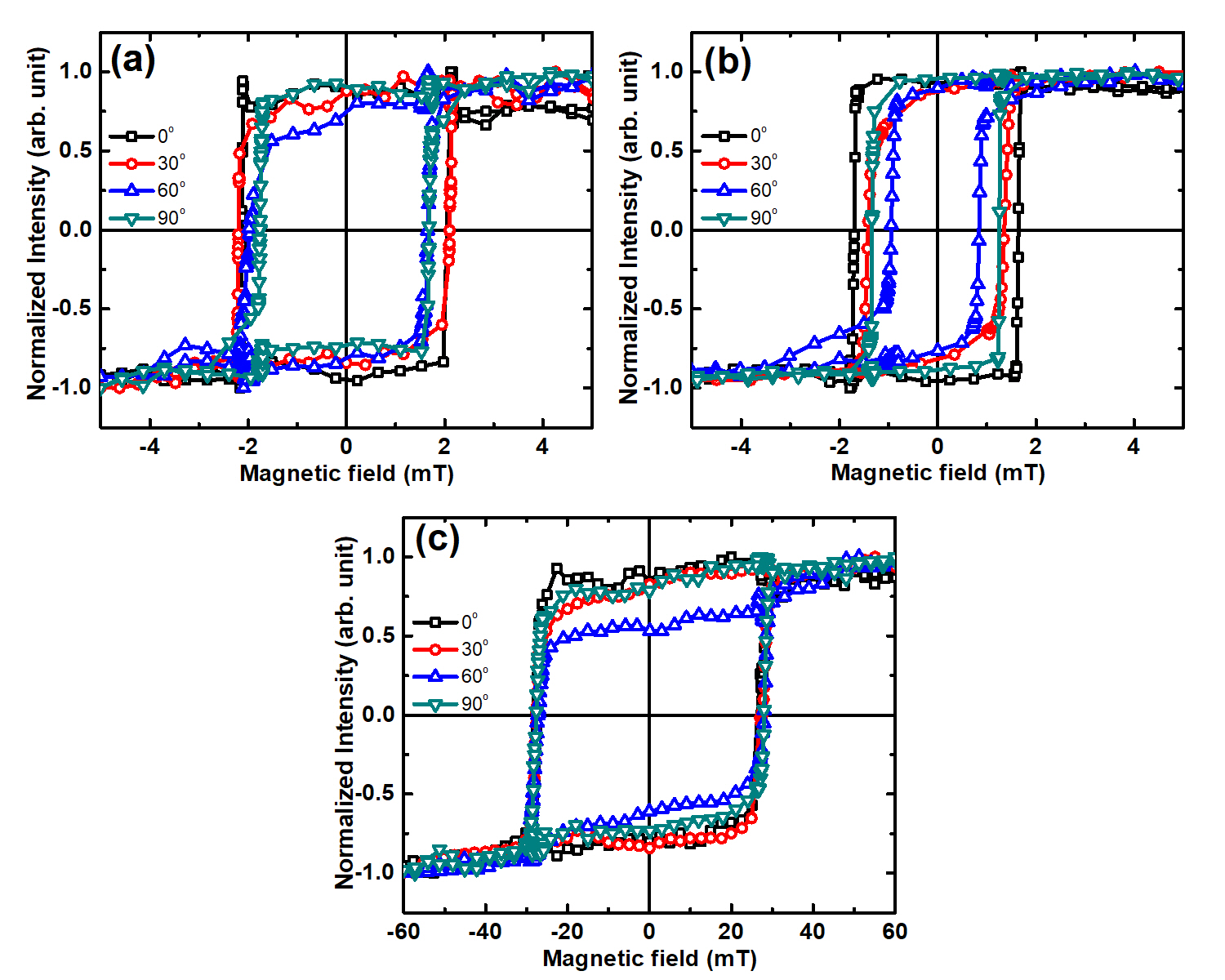}
 \end{center}
   \caption{Hysteresis loops measured at room temperature by longitudinal Kerr microscopy at 0$^{\circ}$, 30$^{\circ}$, 60$^{\circ}$ and 90$^{\circ}$ for samples A - C shown in (a) - (c), respectively..
   \label{Figure1}}
\end{figure*}

$\textit{L}1_0$-FeNi thin films were prepared in a molecular beam epitaxy chamber with base pressure of    $10^{-8}$ Pa consisting of e-beam evaporators (for Fe and Ni) and Knudsen cells (for Cu and Au). First a seed layer of Fe (1 nm) and Au (20 nm) were deposited at 80$^{\circ}$C on MgO (001) substrate followed by Cu (50 nm) layer deposited at 500$^{\circ}$C. It has been reported that Cu and Au layers are prone to form an alloy of Cu$_{3}$Au (001) \cite{mizuguchi2010characterization}. In order to fabricate highly ordered $\textit{L}1_0$-FeNi films, a buffer layer of Au$_{0.06}$Cu$_{0.51}$Ni$_{0.43}$ (50 nm) was deposited on Cu$_{3}$Au layer at 100$^{\circ}$C by co-deposition \cite{kojima2011magnetic}. On top of this buffer layer, FeNi layer was grown by alternative layer deposition (Fe and Ni one after another) or co-deposition (deposition of Fe and Ni simultaneously). Disordered-FeNi film (sample A) was obtained by co-deposition process whereas $\textit{L}1_0$-ordered FeNi films were fabricated by alternate deposition at 100$^{\circ}$C (sample B) and 190$^{\circ}$C (sample C), respectively. Finally, Au capping layer (3 nm) was deposited on top of FeNi layer at 30-40$^{\circ}$C. To study magnetization reversal and magnetic domain structures, we have performed hysteresis measurements along with simultaneous domain imaging using magneto optic Kerr effect (MOKE) based microscope manufactured by Evico Magnetics Ltd. Germany \cite{evico}. Kerr microscopy was performed at room temperature in longitudinal geometry. Angle dependent hysteresis loops were measured by applying the field for 0$^{\circ}$ $\textless \Phi \textless$360$^{\circ}$ at interval of 10$^{\circ}$ where $\Phi$ is the angle between the easy axis and the direction of applied magnetic field. Further, magnetization dynamics of the $\textit{L}1_0$-FeNi thin films has been studied by ferromagnetic resonance (FMR) spectroscopy technique in a flip-chip manner using NanOsc Instrument Phase FMR \cite{fmr}. Frequency range of the RF signal used in this experiment was between 5 and 17 GHz. To understand the symmetry of anisotropy and quantify the anisotropy energy density, angle dependent FMR measurement has been performed by applying the magnetic field in the sample plane with a fixed frequency of 7 GHz. 

Magnetic hysteresis loops measured using Kerr microscopy for samples A-C are shown in figure 1 for various values of Φ = 0$^{\circ}$, 30$^{\circ}$, 60$^{\circ}$ and 90$^{\circ}$. Square shaped hysteresis loops have been observed for all three samples for all the angles (0$^{\circ}$ $\textless\Phi \textless$360$^{\circ}$). This indicates that the magnetization reversal occurs via nucleation and subsequent domain wall motion \cite{mallick2014magnetic}. It is observed that the coercivity varies significantly among the samples e.g. \textit{H$_{C}$} of samples A, B and C are ~ 2.2, 1.7 and 25 mT, respectively. \textit{K$_{u}$} of these samples are -1.05$\times$10$^{6}$ erg cm$^{-3}$(A), 3.47$\times$10$^{6}$ erg cm$^{-3}$(B), and 4.86$\times$10$^{6}$ erg cm$^{-3}$(C).

\begin{figure*}[htb]
   \begin{center}
   \includegraphics[scale=0.5]{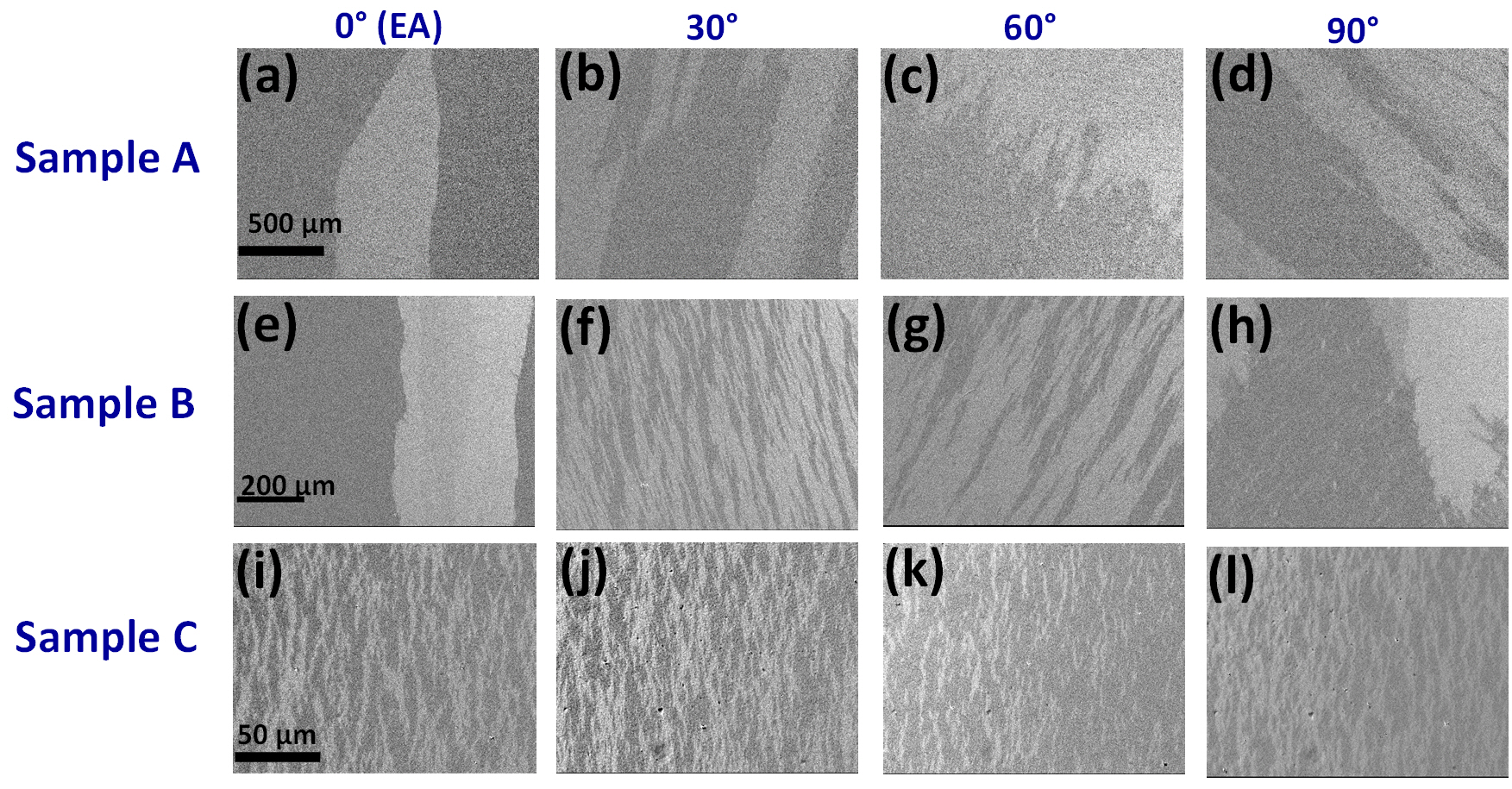}
 \end{center}
   \caption{Domain images captured for samples A, B and C at angles, 0$^{\circ}$, 30$^{\circ}$, 60$^{\circ}$ and 90$^{\circ}$ near to coercivity are shown. Scale bars are shown on the domain images for each sample separately which are valid for the images recorded at other angles for those respective samples.
   \label{Figure2}}
\end{figure*}

Domain images captured near to coercive field \textit{H$_{C}$} for all three samples at 0$^{\circ}$, 30$^{\circ}$, 60$^{\circ}$ and 90$^{\circ}$ are displayed in figure 2. The black and gray contrast in the domain images represent positive and negative magnetized states, respectively. From the domain images it can be concluded that magnetization reversal occurs via domain wall motion. Sample A (figure 2 a to d) exhibits large well-defined stripe domains which indicates the presence of weak magnetic anisotropy. These stripe domains are found to be tilted for e.g. $\Phi$ = 30$^{\circ}$ and 60$^{\circ}$ (figure 2 b and c). For $\Phi$ = 90$^{\circ}$ (figure 2d) branched domains are observed. In addition to 180$^{\circ}$ domain wall, sample B (figure 2 e to h) shows 90$^{\circ}$ domain walls at certain angles of measurement (figure S1). Sample C (figure 2 i to l) shows narrow branched domains and are found to be independent of $\Phi$ , which indicates that the sample C is magnetically isotropic in nature. By comparing the domain images at any particular $\Phi$ among the samples, it is observed that the size (i.e. width) of the domain decreases with increasing anisotropy strength.

In the following we have investigated the magnetization dynamics of the $\textit{L}1_0$–FeNi films by FMR. Measured FMR spectra (open symbol) for samples A and C at selective frequencies are shown in figure 3 (a) and (b), respectively. Resonance field (H$_{res}$) and line width ($\Delta H$) were extracted by fitting of Lorentzian shape function (equation 1) having anti-symmetric (first term) and symmetric components (second term) to the obtained FMR derivative signal \cite{singh2017study};

\begin{equation}
\begin{split}
S_{21}&=K_{1}\frac{4\Delta H(H-H_{res})}{[4(H-H_{res})^2+(\Delta H)^2]^2} \\
&-K_{2}\frac{(\Delta H)^2-4(H-H_{res})^2}{[4(H-H_{res})^2+(\Delta H)^2]^2} \\
&+ (slope H) + Offset
\end{split}
\end{equation}

where \textit{S$_{21}$} is transmission signal, \textit{K$_{1}$} and \textit{K$_{2}$} are coefficient of anti-symmetric and symmetric component, respectively, \textit{slope H} is drift value in amplitude of the signal. Frequency (\textit{f}) dependence of \textit{H$_{res}$} and \textit{$\Delta$H} are plotted in figure 3 (a) and (b), respectively. From the \textit{f} vs. \textit{H$_{res}$} plot, parameters such as Lande g-factor, effective demagnetization field (4\textit{$\pi$M$_{eff}$}), anisotropy field \textit{H$_{K}$} were extracted by fitting it to Kittel resonance condition \cite{kittel1948theory};

\begin{equation}
f=\frac{\gamma}{2\pi} \sqrt{(H_{K}+H_{res})(H_{K}+H_{res}+4\pi M_{eff})}
\end{equation}

where $\gamma = g \mu_{B}/\hbar$ is gyromagnetic ratio, $\mu_{B}$ as Bohr magneton, $\hbar$ as reduced Planck’s constant. The values of damping constant $\alpha$ for all three samples were extracted by using the equation \cite{singh2017study,kittel1948theory,heinrich1985fmr};

\begin{equation}
\Delta H = \Delta H_{0} + \frac{4 \pi \alpha f}{\gamma}
\end{equation}

\begin{figure*}[htb]
	\begin{center}
		\includegraphics[scale=0.5]{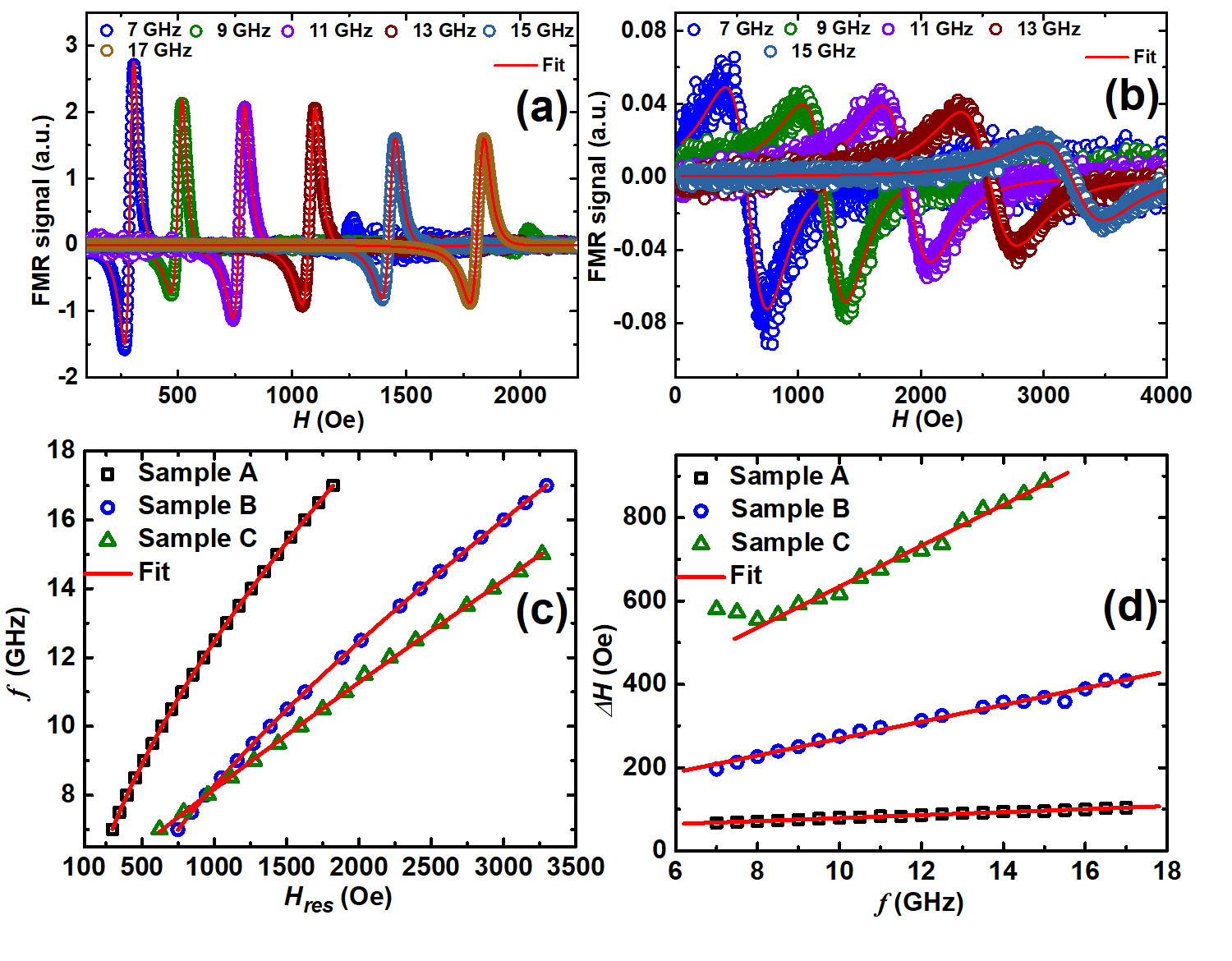}
	\end{center}
	\caption{FMR spectra of (a) sample A and (b) sample C measured at frequency 5 – 17 GHz. The solid lines in (a) and (b) are fitted with equation 1. (c) \textit{f}  vs. \textit{H$_{res}$} and (d) $\Delta H$ vs. \textit{f}  plots for samples A – C extracted from the FMR frequency dependent spectra. The lines in (c) and (d) are the best fits to the equations (2) and (3), respectively.
		\label{Figure3}}
\end{figure*}

\begin{table*}[]
	\caption{\label{tab:table1}The Parameters extracted from the fitting of FMR experimental data (Figure 3) of all three samples using the equations (2) and (3)}
	\begin{ruledtabular}
	\begin{tabular}{cccccc}
		\textbf{Sample} & \textbf{H$_{K}$(Oe)}   & \textbf{$4\pi M_{eff}$} & \textbf{g-factor} & \textbf{$\alpha$} & \textbf{$\Delta$ H$_{0}$(Oe)} \\ \hline
		A               & 49.41$\pm$0.82    & 18279$\pm$365       & 1.97$\pm$0.01        & 0.0049$\pm$0.0001  & 42.85$\pm$0.65    \\
		B               & -52.39$\pm$4.64   & 8900$\pm$193        & 1.93$\pm$0.01        & 0.0277$\pm$0.0007  & 66.89$\pm$7.13    \\
		C               & 680.29$\pm$106.88 & 3509$\pm$106        & 1.98$\pm$0.01        & 0.0680$\pm$0.0019  & 143.67$\pm$16.20  \\
	\end{tabular}
	\end{ruledtabular}
\end{table*}

where $\Delta H_{0}$ is called as inhomogeneous line width broadening. Values of all the fitting parameters obtained by using equations (1) and (2) are given in table 1. With increase in anisotropy strength of the samples, $4\pi M_{eff}$ decreases whereas $\alpha$ increases. Sample A exhibits lowest value of $\alpha$, $4.9 \times 10^{-3}$ which is the same order with normal FeNi alloy thin film ($\alpha = 1.2 \times 10 ^{-3}$) \cite{zhu2018static}. The value of inhomogeneous line width broadening $\Delta H_{0}$, which depends on the quality of the thin film \cite{singh2017study}, is highest for sample C and lowest for sample A.

\begin{figure*}[htb]
	\begin{center}
		\includegraphics[scale=0.4]{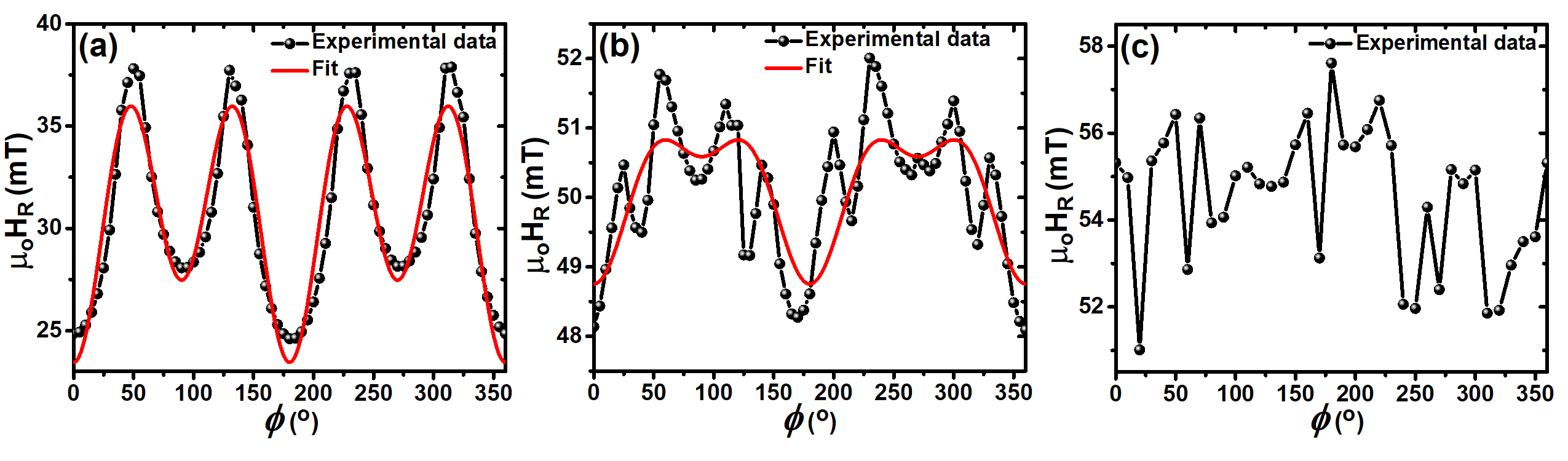}
	\end{center}
	\caption{Anisotropy symmetry plot and the fits for (a) sample A, (b) sample B and (c) sample C measured using FMR by keeping the frequency fixed at 7 GHz.  The red solid lines are the best fits to equation (4). 
		\label{Figure4}}
\end{figure*}

Apart from frequency dependent FMR, angle dependent FMR measurements at a fixed frequency of 7 GHz were performed to analyze the anisotropy energy density as well as symmetry of all three samples. Figure 4 shows the angle dependent \textit{H$_{res}$} plot for all three samples. Samples A and B clearly show the presence of mixed cubic and uniaxial anisotropies with minima in \textit{H$_{res}$} approximately at angles 0$^{\circ}$, 90$^{\circ}$, 180$^{\circ}$ and 270$^{\circ}$. From the angle dependent \textit{H$_{res}$} plot, the value of uniaxial and cubic anisotropy energy constants have been estimated by fitting those data to solution of LLG equation that includes the cubic and uniaxial anisotropies and it is written as \cite{pan2017role};

\begin{equation}
\begin{split}
	f&= \frac{\gamma}{2\pi} \left[ H + \frac{2K_{2}}{M_{s}} cos2\phi - \frac{4K_{4}}{M_{s}}cos4\phi \right]\\ 
	&\times \left[ H + 4\pi M_{s} +\frac{2K_{2}}{M_{s}} cos2\phi - \frac{K_{4}}{M_{s}}(3+cos4\phi) \right]^{1/2} 
\end{split}
\end{equation}

\begin{table}[]
	\caption{\label{tab:table2}The parameters extracted from the fitting of angle dependent FMR experimental data of two samples using the equation (4)}
	\begin{ruledtabular}
		\begin{tabular}{cccc}
			
			\textbf{Samples} & \textbf{$M_{s}$(emu cm$^{-3}$)}      & \textbf{$K_{2}$(erg cm$^{-3}$)}     & \textbf{$K_{4}$(erg cm$^{-3}$)}     \\ \hline
			\textbf{A}       & 1645             & 1.66$\times$10$^{4}$        & -2.16$\times$10$^{4}$       \\
			\textbf{B}       & 1026.8           & 4.8$\times$10$^{3}$         & -1.2$\times$10$^{3}$        \\
			\textbf{C}       & \multicolumn{3}{c}{Magnetically isotropic behaviour} \\ 
		\end{tabular}
	\end{ruledtabular}
\end{table}

where, \textit{K$_{2}$} and \textit{K$_{4}$} are cubic and uniaxial anisotropy energy density constants, respectively, \textit{M$_{S}$} is saturation magnetization, \textit{H} is the applied magnetic field, \textit{$\Phi$} is the angle between applied field direction and easy axis of the sample. The angle dependence of \textit{H$_{res}$} is fitted to equation 4. From the fit, value of \textit{M$_{S}$}, \textit{K$_{2}$} and \textit{K$_{4}$} have been extracted and are shown in table 2. It has been found that both cubic and uniaxial anisotropy energy constants of sample A are greater than that of sample B by one order of magnitude. Further in sample A the ratio between the cubic to uniaxial anisotropy is about ~0.75 whereas for sample B the ratio is 4. Therefore for sample B the cubic anisotropy is four times stronger than the uniaxial anisotropy. This is the reason of occurrence of 90$^{\circ}$ domain walls for sample B as shown in Fig. S1\cite{mallik2014interplay}. In comparison to samples A and B, sample C shows isotropic behavior which is in consistent with the results obtained from Kerr microscopy.

$\textit{L}1_0$-FeNi thin films (samples A, B and C) show different strength of anisotropy, which were fabricated by different deposition processes in a MBE system. Domain images reveals that magnetization reversal for all three samples occur via nucleation and subsequent domain wall motion. We have observed lowest values of $\alpha$ $\sim$ $4.9\times10^{-3}$ for the sample A which shows moderate magnetic anisotropy. Angle dependent FMR measurement shows the mixed cubic and uniaxial anisotropy in samples A and B, while sample C exhibits magnetically isotropic behavior. Our work demonstrates that variable anisotropic $\textit{L}1_0$-FeNi thin films can be fabricated in which the damping constant and magnetization reversal can be tuned. These results may be useful for future spintronics based applications.

Acknowledgements:
We acknowledge the financial support by department of atomic energy (DAE), Department of Science and Technology (DST-SERB) of Govt. of India,DST-Nanomission (SR/NM/NS-1018/2016(G)) and DST, government of India for INSPIRE fellowship.

\section*{References}



%

\section*{Supplementary Material}

\begin{figure}[htb]
	\begin{center}
		\includegraphics[scale=0.6]{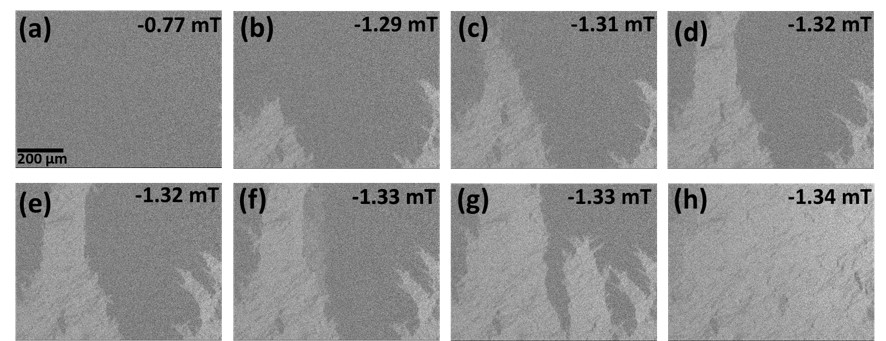}
	\end{center}
	\caption{Magnetic domain images for sample B at various fields close to the reversal field. It shows that the magnetization reversal is happening through 90$^{\circ}$ domain wall in sample B.
		\label{FigureS1}}
\end{figure}

\end{document}